\documentstyle [aps]{revtex}
 
\newcommand{\la}[1]{\label{#1}}

\newcommand{\A}{{\bf A}}

\renewcommand{\d}{{\bf d}}

\newcommand{\n}{{\bf n}}

\newcommand{\be}{\begin{equation}}
\newcommand{\ee}{\end{equation}}
\newcommand{\ba}{\begin{eqnarray}}
\newcommand{\ea}{\end{eqnarray}}
\newcommand{\bastar}{\begin{eqnarray*}}
\newcommand{\eastar}{\end{eqnarray*}}

\begin{document}
 
\title  {Monopole condensates in SU($N$) Yang-Mills Theory
}

\author{Vipul Periwal} 

\address{Department of 
Physics,
Princeton University,
Princeton, New Jersey 08544}
\def\dd{\hbox{d}}
\def\tr{\hbox{tr}}\def\Tr{\hbox{Tr}}
\def\eee#1{{\rm e}^{{#1}}}
\def\eE#1{{\rm e}^{{#1}}}
\def\refe#1{eq.~\ref{#1}}
%\date{\today}
\maketitle

\begin{abstract}
Faddeev and Niemi have proposed a reformulation of SU(2) Yang-Mills theory in 
terms of new variables, 
appropriate for describing the theory in its infrared limit
based on the intuitive picture of colour confinement due to monopole 
condensation.  I generalize their proposal (with some 
differences) to SU($N$) Yang-Mills 
theory.  The natural variables are $N-1$ mutually commuting traceless
$N\times N$ Hermitian matrices, an element of the maximal torus 
defined by these commuting matrices, $N-1$ Abelian gauge fields for the 
maximal torus gauge group, and an invariant symmetric two-index tensor on the 
tangent space of the maximal torus, adding up to the requisite  
2$(N^{2}-1)$ physical degrees of freedom. 

\end{abstract}

%\narrowtext
\widetext
\bigskip

Twenty-five years after the discovery of asymptotic freedom\cite{asymp}, 
there is still no quantitative theoretical understanding of colour confinement
in Yang-Mills theories.   The qualitative picture of 
confinement\cite{thooft}\ due to
monopole condensation has yet to be given a quantitative 
justification starting from the fundamental Yang-Mills action.

Faddeev and Niemi\cite{fad}\ have recently proposed 
that even though 
the gauge field $A_\mu$ is the proper order parameter 
for describing the theory in its ultraviolet limit, in 
the infrared limit with monopole condensation
some other order parameter may be more appropriate.  They argue that
in the infrared limit, for the SU(2) gauge theory, 
the appropriate order parameter 
is a three component vector $n^a(x) \ (a = 1,2,3)$
with unit length $\n \cdot \n = 1$ and
classical action \cite{fadde}
\be
S \ = \ \int \dd x \left[ m^2 (\partial_\mu \n)^2 \ + \
\frac{1}{e^2} ( \n , \d\n \times \d\n)^2\right] 
\label{fad1}
\ee
with $m$ a mass scale and $e$ a dimensionless 
coupling constant. This is the {unique} 
local and Lorentz-invariant action for the unit 
vector $\n$ which is at most quadratic in time 
derivatives so that it admits a Hamiltonian 
interpretation, and involves { all} infra-red 
relevant (or marginal) terms with this property.

Following some ideas of Cho\cite{cho}, 
they derived  \refe{fad1}\ by parametrizing a generic four 
dimensional SU(2) connection 
\be
\A_\mu \ \equiv \ C_\mu \n \ + \ \d\n \times \n \ + \ \rho  
\d \n \ + \ \sigma \d\n \times \n ,
\la{wu5}
\ee
with an Abelian gauge field $C_{\mu}$  and scalar fields $\rho,\sigma$.  
They motivated this decomposition by considering 
the gauge transformation properties of $A_{\mu}$ under
gauge transformations generated by $\n,$ regarded as an element of 
the SU(2) Lie algebra.  They noted that
while $A$ has 2$\cdot$ dim(SU(2))  physical degrees of freedom in four 
dimensions, 
$C$ has two, $\n$ has 
two, and $\rho,\sigma$ provide the requisite remaining  two.  In particular, an 
amusing suggestion\cite{fad}\ was that $\phi = \rho + i \sigma$ 
transforms as a complex scalar field under U(1) gauge transformations
generated by $\alpha \cdot \n.$   Note that this
decomposition is specific to gauge theories in four dimensions.  The 
effective description (\refe{fad1}) is therefore 
dimension-specific---in other words, the action given in \refe{fad1}\ 
can be defined in any number of dimensions, but can  describe 
low-energy Yang-Mills theory only in four dimensions.
 
Much intuition for the behaviour of gauge theories is based on the
large $N$ limit.  It is therefore of interest to see how the
Faddeev--Niemi decomposition generalizes for arbitrary gauge groups.  
For instance, what replaces the complex
scalar field in the Faddeev--Niemi decomposition? Which degrees of 
freedom dominate in the large $N$ limit? 
Cho already considered the SU(3) case\cite{cho2} to some extent, but 
his arguments seem to involve an extension of Yang-Mills theory, and 
are therefore not directly related to the approach taken by Faddeev and 
Niemi\cite{fad}.
I shall present below the general case of SU($N$), which turns out to
have considerable geometric intricacy compared to SU(2).  It appears 
as somewhat of a miracle that the general parametrization
given below (\refe{complete}) actually 
leads to a correct matching of degrees of freedom for any $N $ 
(\refe{counting}).  The 
final form will not exactly match the properties remarked on by
Faddeev and Niemi\cite{fad}\ for SU(2), in that the variables for
arbitrary $N$ do not appear to admit a straightforward Higgs 
interpretation.

\def\DDD{{\cal D}}
For SU($N$), it is  obvious that $\n(x)$ is replaced by $N-1$ commuting
traceless Hermitian $N\times N$ matrices, $H_{I}(x),$
orthonormal with respect to  the trace, $\tr H_{I}H_{J}= 
\delta_{IJ}.$  How many parameters does it take to specify such a 
set of matrices?  The simplest count is to consider that the dimension
of the space of maximal tori of SU($N$) is $N^{2}-N,$ and we further 
need to specify an orthonormal frame in the tangent space of the
maximal torus, which requires an SO($N$-1) transformation,
giving $N^{2}-N+(N-2)(N-1)/2$
degrees of freedom.  We thus have an $N-1$ dimensional bundle of
commuting, trace-orthonormal, traceless Hermitian matrices over
spacetime.  On this bundle, there is a natural spin connection,
$\DDD_{\mu}H_{I}\equiv \partial_{\mu}H_{I} + \omega_{\mu I}{}^{J}H_{J},$
with $\omega$
specified completely by 
\be
\tr H_{I}\DDD_{\mu}H_{J} = 0 \Rightarrow \omega_{\mu IJ} = - 
\tr H_{J}\partial_{\mu}H_{I}.
\la{spinit}
\ee
The $I,J$ indices on $\omega$ are raised and lowered with the
identity matrix. $\omega_{\mu IJ}$ is antisymmetric 
in $I,J$ since $\partial_{\mu} \tr H_{I}H_{J} =0.$ 

The point of introducing the spin connection (\refe{spinit})
is that  in terms of $\{H_{I}\},$ and $\DDD_{\mu},$
I can write
\be
A_{\mu} \equiv C_{\mu}^{I}H_{I} +\phi^{I}\DDD_{\mu}H_{I} +
i\beta^{IJ}\left[H_{I}, \DDD_{\mu}H_{J}\right] + {\rm nonlinear\ terms},
\la{incomplete}
\ee
with $C_{\mu}^{I} \equiv \tr A_{\mu}H_{I}.$
Notice that $\left[H_{I}, \DDD_{\mu}H_{J}\right]$ is symmetric in
$I$ and $J,$ since $\partial_{\mu}\left[H_{I}, H_{J}\right]=0.$  
Thus
$\DDD_{\mu}H_{J}$ and $\left[H_{I}, \DDD_{\mu}H_{K}\right]$ are 
linearly independent.  For SU($N$), $N>2,$ there is no way to
express 
further commutators of the form $\left[H_{I},\left[H_{K}, \dots 
\DDD_{\mu}H_{J}\right]\dots \right]$ in terms of these two types of 
terms.  As is obvious, such linearly independent
higher commutators will automatically be 
generated by the action of gauge transformations.  As I shall show, 
it is possible to include terms nonlinear in $\phi$ in such a way that
gauge transformations corresponding to the maximal torus of SU($N$) 
defined by $H_{I}$ 
preserve the complete form of $A_{\mu}.$

Before considering the form of the requisite nonlinear terms,
we should count degrees of freedom introduced in \refe{incomplete}.
Besides the degrees of freedom in $H_{I},$ we get $2(N-1)$ from the $N-1$
Abelian gauge fields $C_{\mu}^{I},$ $N-1$ from the scalar fields
$\phi^{I},$ and $N(N-1)/2$ from the scalar fields 
$\beta^{IJ}.$  Therefore 
\be
2(N^{2}-1)= \left(N^{2}-N +(N-2)(N-1)/2\right)  + 2(N-1) +
N-1 + N(N-1)/2 ,
\la{counting}
\ee
which is exactly correct.  In particular, if we added further degrees 
of freedom associated with multiple commutators, we would have too
many degrees of freedom.

The appearance of the spin connection $\DDD_{\mu} $ leads to several 
significant differences  from the SU(2) case---obviously, the
spin connection reduces to the ordinary derivative for the case of a
real line bundle.  Consider a gauge transformation generated by
$\chi \equiv\alpha^{I}(x)H_{I}(x).$  Such gauge transformations are
particularly significant in our rewriting of the gauge field, since 
they lie in  the  ${\rm U}(1)^{N-1}$ maximal torus subgroup of SU($N$)
gauge group 
defined by $H_{I}.$  We have
\be 
\delta A_{\mu} = \partial_{\mu}\chi + i[A_{\mu},\chi],
\la{gaugetra}
\ee
which we compare to 
\be 
\delta A_{\mu} = \delta C_{\mu}^{I}H_{I} +\delta \phi^{I}\DDD_{\mu}H_{I} +
i\delta \beta^{IJ}\left[H_{I}, \DDD_{\mu}H_{J}\right] + \delta({\rm 
nonlinear\ terms}).
\la{variation}
\ee
Defining the action of the spin connection on $\alpha^{I}$
in the obvious manner, $(\DDD_{\mu}\alpha)^{I} \equiv  
\partial_{\mu}\alpha^{I} - \alpha^{J}\omega_{\mu J}{}^{I},$ we have
\be
\partial_{\mu}\chi = (\DDD_{\mu}\alpha)^{I}H_{I} + \alpha^{I}
(\DDD_{\mu}H)_{I},
\ee
so that 
the Abelian gauge transformations of $C_{\mu}^{I}$ involve the spin connection:
\be 
\delta C_{\mu}^{I} = (\DDD_{\mu}\alpha)^{I},
\ee
which is natural from the perspective  of monopole moduli.

The behaviour of $\phi$ and $\beta$ under such gauge 
transformations, \refe{gaugetra}, is  
\be 
\delta \phi^{I}  = \alpha^{I},
\la{phitra}
\ee
with 
\be
\delta \beta^{IJ} = -{1\over 2} \left(\phi^{I}\alpha^{J} + \phi^{J}\alpha^{I}
\right).
\la{betatra}
\ee
However, we clearly need to find appropriate nonlinear terms 
to complete \refe{incomplete}, since not all terms in \refe{gaugetra} 
have been correlated with terms in \refe{variation}.  The first step in
finding these nonlinear terms is to note that \refe{betatra} shows that
if we shift $\beta$ by ${1\over 2}\phi^{I}\phi^{J},$ then the redefined
$\beta$ does not transform at all.  Further, \refe{phitra} suggests 
that we should rewrite \refe{incomplete} in a manner that  makes 
this translation  symmetry manifest.

\def\ap#1{{\rm Ad}_{{#1}}}
Define ${\rm Ad}_{\phi}(X) \equiv \eE{-i\phi^{I}H_{I}}X\eE{i\phi^{I}H_{I}}.$
Consider then 
\be
A_{\mu} \equiv C_{\mu}^{I}H_{I} + 
 \ap\phi({E_{\mu}}) - E_{\mu} + i\beta^{IJ}\left[H_{I}, 
\ap\phi({\DDD_{\mu}H_{J}})\right]  ,
\la{complete}
\ee
where $E_{\mu}$ is implicitly defined by
$-i[H_{I},E_{\mu}]\equiv\DDD_{\mu}H_{I}.$
(Notice that \refe{complete} can be computed without ever computing 
$E_{\mu}$ explicitly, so the introduction of $E_{\mu}$ is solely to 
make \refe{complete} transparent.  However, the intuition for writing
\refe{complete} in terms of $E_{\mu}$ comes from observing that 
$[H_{I},\DDD_{\mu}H_{J}]$ is symmetric in $I$ and $J,$ and that 
$\tr H_{I}\DDD_{\mu}H_{J}=0.$)  
An alternate way of writing \refe{complete} is as
\be 
A_{\mu} + E_{\mu}= \ap\phi(C_{\mu}^{I}H_{I}  + i\beta^{IJ}\left[H_{I}, 
 {\DDD_{\mu}H_{J}} \right] + 
 {E_{\mu}})   .
\ee
We deduce  from \refe{complete}
that $\phi^{I}$ should be regarded as compact variables taking values on the
maximal torus defined by $H_{I}$---this should be compared to the 
variables introduced in \cite{fad}.

Now, under an infinitesimal 
gauge transformation with
parameter $\alpha\cdot H\equiv \alpha^{I}H_{I},$ we have
\be
A_{\mu}(C_{\mu},\phi,\beta) \rightarrow A'_{\mu} = 
A_{\mu}(C_{\mu}+\DDD_{\mu}\alpha,\phi+\alpha,\beta).
\la{gauge}
\ee
This remarkably simple form, \refe{gauge}, implies that 
under a  gauge transformation with parameter 
$\alpha^{I}=-\phi^{I},$
the transformed gauge connection takes the form
\be
A'_{\mu} = (C_{\mu}-\DDD_{\mu}\phi)\cdot H + 
i\beta^{IJ}\left[H_{I}, \DDD_{\mu}H_{J}\right] + \hbox{terms nonlinear 
in $\phi$},
\la{aprime}
\ee
which makes the provenance of the 
physical degrees of freedom more transparent.
The nonlinear terms  are 
\be
E_{\mu} - \int_{0}^{1}\dd s \ap{-s\phi}(\phi\cdot \DDD_{\mu}H)
- \ap{-\phi}(E_{\mu}) .
\ee
There is, 
of course, no gauge symmetry evident in \refe{aprime}, since the 
field-dependent gauge transformation has fixed it completely, 
leading to a gauge in which $A'$ contains no terms
linear in $\phi,$ other than the derivative term appearing with $C_{\mu}.$

What is the analogue of \refe{fad1}\ for SU($N$)?  Computing the 
curvature of the SU($N$) connection, as parametrized in 
\refe{complete}, and inserting into the Yang-Mills action, 
neglecting terms involving 
$\phi$ and $C$ for the nonce, with   
$\beta^{IJ}(x)=\lambda \delta^{IJ}={\rm 
const.},$ we find 
\be
S \ = - \int \dd x \  
\tr \left(2\lambda 
\left[\partial_{\mu}H_{I},\partial_{\nu}H^{I}\right] -\lambda^{2}
\left[\left[H_{I}, \partial_{\mu}H^{I}\right],\left[H_{K}, 
\partial_{\nu}H^{K}\right]\right]\right)^{2}\ .
\label{lowaction}
\ee
Notice that $\lambda=0$ is a singular point, which is entirely 
appropriate  when we 
recall the motivation of Cho\cite{cho,cho2}, and Faddeev and 
Niemi\cite{fad}---$\beta=0$ amounts essentially to enforcing that the 
monopole condensate does not contribute to the   gauge potential in
\refe{complete}.  It should also be noted that local frame rotations 
$H_{I}(x) \rightarrow O^{J}_{I}(x) H_{J}(x)$ are {\it not} a symmetry 
of the action, but physical degrees of freedom (which have been 
included in \refe{counting}).
While this action (\refe{lowaction})
admits a Hamiltonian interpretation, and also possesses
infrared relevant deformations that would not appear in the
Yang-Mills action, for example,
\be
\Delta S \ =  \int \dd x \left\{ m_{1}^{2}\ \tr \ \partial_{\mu}
H\cdot \partial^{\mu}H 
+ m_{2}^{2}\ \tr \left[H_{I}, \partial_{\mu}H_{K}\right]
\left[H^{I}, \partial^{\mu}H^{K}\right]\right\},
\ee
it is absolutely crucial to keep in mind that the $\phi^{I}$ variables
live on a maximal torus, related by a gauge invariance to
the $C_{\mu}$ fields, so it is not correct to neglect
these fields.  A relevant deformation of the form
\be
 S' \ =  \int \dd x \ M^{2} (C_{\mu}-D_{\mu}\phi)^{2}
\ee
preserves (Abelian) gauge invariance (\refe{gauge}), and would serve to
render the  the gauge field massive (and hence 
eliminate it from the low energy dynamics of $H$ if $M$
were large enough relative to $m_{i}$).  As required, the 
$\phi^{I}$ fields carry exactly enough degrees of freedom to provide 
all three polarizations of  massive vector bosons, when combined with
$C_{\mu}^{I}.$   
In view of the fact that we are attempting a 
description of  
the confining behaviour of Yang-Mills theory, it is extremely 
important that {\it all} excitations have the {\it possibility} of 
acquiring a mass gap.   

I do not go into more detail here  regarding \refe{lowaction}\ or its 
relevant perturbations because this is properly a matter of 
detailed calculation.  The entirely reasonable claim\cite{fad} 
is that these relevant operators will appear when one integrates out
higher momentum modes in a Wilsonian approach.  In fact, one  
expects that all the 
relative coefficients of all the relevant perturbations should be 
calculable in such an approach. (There are   
other relevant operators besides those mentioned above.)
I hope to return to this  elsewhere.

Comparing \refe{complete} to the form (\refe{wu5}) 
suggested by Faddeev and Niemi\cite{fad},  
$\rho = (1-\beta)\sin\phi,$ and $\sigma= (1-\beta)\cos\phi.$
$\beta$ does not 
transform under the Abelian gauge symmetry, hence an expectation
value for $\beta$ does not give rise in an obvious manner to a Higgs-like phase.   
However, since $(1-\beta)^2 = \rho^2+\sigma^2,$   roughly speaking 
$1-\beta$ is the magnitude of the Higgs field.   So 
$\beta =1$ is the phase with vanishing
Higgs expectation value, which is also the value at which the Lagrangian 
for $H$ simplifies.
On the other hand, when $\beta=0,$ the Lagrangian for the 
$H$ variables vanishes, which one
could call a Higgs phase, but is so singular that one 
has to be a trifle more careful in 
determining exactly what happens in the dynamics.

In conclusion, I have found a simple generalization of the  
decomposition suggested by Faddeev and Niemi\cite{fad}\
to arbitrary rank gauge groups.  The natural variables 
in the SU($N$) case are  $H_{I}$, a maximal Abelian subalgebra of the Lie 
algebra, $\phi^{I},$ which are essentially 
coordinates on the maximal torus defined by 
$H_{I},$ $\beta^{IJ},$ a symmetric two-index tensor field on the tangent 
space of this maximal torus, and $C_{\mu}^{I},$ 
$N-1$ Abelian gauge fields associated 
with $H_{I}.$  Rather miraculously, the physical degrees of freedom in
the SU($N$) gauge field match the physical degrees of freedom in
these fields, \refe{counting}.  The $H_{I}$ and $\beta^{IJ}$ fields have
$O(N^{2})$ degrees of freedom and may be expected to dominate the dynamics in 
the large $N$ limit.  From a canonical point of view, this change of 
variables is perfectly benign in the classical Hamiltonian.  From the
quantum point of view, the construction of a confinement vacuum may be
simpler in the new variables. I have not addressed any 
issues of electric-magnetic duality for the simple reason that
it is not at all clear that the change of variables(\refe{complete}) 
{\it should} make both the confinement and the Higgs phases of Yang-Mills
theory transparent\cite{stan}.  
Indeed, one might expect that an electric version 
of \refe{complete}\ would be a more appropriate starting point for
describing a possible Higgs phase.  A low-energy effective action, 
governing the 
dynamics of the $H_{I}$ field  is calculable starting from 
the Yang-Mills action and \refe{complete}, 
and should possess knot-like solutions
analogous to those found by Faddeev and Niemi\cite{nature}.  Further, 
it would be interesting to compare this action to \cite{bard}.

{\bf Aknowledgements} I am grateful to Wati Taylor and Peter Schupp for
helpful comments.  I am especially grateful to Y.J. Ng in correcting 
my interpretation of \cite{fad}.
This work was supported in part by NSF grant PHY96-00258.

\end{document}